\documentclass[aps,prl,twocolumn,superscriptaddress]{revtex4}
\usepackage[dvips]{color}
\usepackage{colordvi}
\usepackage{graphicx}

\begin{document}

%\preprint{}
\title{Magnetic Excitations and their energy change available to Superconducting
Condensation in Optimally Doped YBa$_2$Cu$_3$O$_{6.95}$}

\author{Hyungje Woo}
\affiliation{
Department of Physics and Astronomy, The University of Tennessee, Knoxville, Tennessee 37996-1200, USA
}
\affiliation{
Center for Neutron Scattering, Oak Ridge National Laboratory, Oak Ridge, Tennessee 37831, USA}
\author{Pengcheng Dai}
\affiliation{
Department of Physics and Astronomy, The University of Tennessee, Knoxville, Tennessee 37996-1200, USA
}
\affiliation{
Center for Neutron Scattering, Oak Ridge National Laboratory, Oak Ridge, Tennessee 37831, USA}
\author{S. M. Hayden}
\affiliation{
H. H. Wills Physics Laboratory, University of
Bristol, Bristol BS8 1TL, UK
}
\author{H. A. Mook}
\affiliation{
Center for Neutron Scattering, Oak Ridge National Laboratory, Oak Ridge, Tennessee 37831, USA}
\author{T.~Dahm}
\affiliation{\sl Institut f\"ur Theoretische Physik,
         Universit\"at T\"ubingen,
         Auf der Morgenstelle 14, D-72076 T\"ubingen,
         Germany}
\author{D. J. Scalapino}
\affiliation{ Department of Physics, University of California,
Santa Barbara, California 93106, USA }
\author{T. G. Perring}
\affiliation{
ISIS
Facility,Rutherford Appleton Laboratory,Chilton, Didcot, Oxon,
OX11 0QX, UK}
\author{F. Do$\rm\breve{g}$an}
\affiliation{
Department of Ceramic Engineering, University of
Missouri-Rolla, Rolla, Missouri 65409-0330, USA
}

\maketitle

{\bf
Understanding the magnetic excitations in high-transition
temperature (high-$T_c$) copper oxides is important
because they may mediate the electron pairing for superconductivity
\cite{Sca95,CPS03}. By determining
the wavevector ({\bf Q}) and energy ($\hbar\omega$)
dependence of the magnetic excitations,
one can calculate the change in the exchange energy available to
the superconducting condensation energy \cite{SW98,DZ98,maier}.
For the high-$T_c$ superconductor YBa$_2$Cu$_3$O$_{6+x}$,
the most prominent feature in the magnetic excitations is the
resonance \cite{mignod,mook,mook98,fong,dai01,stock,hayden}. Although
the resonance has been suggested to contribute a major part of the
superconducting condensation \cite{DZ98,dai99}, the accuracy of
such an estimation has been in doubt because the resonance is only a small portion
of the total magnetic scattering \cite{hayden,dai99,kivelson}.
Here we report 
an extensive
mapping of magnetic excitations
for YBa$_2$Cu$_3$O$_{6.95}$ ($T_c\approx 93$ K).
Using the absolute intensity measurements of
the full spectra, we estimate the change in the magnetic exchange energy between the
normal and superconducting states and find it to be about 15 times larger than the superconducting condensation energy \cite{loram,Lortz}.
Our results thus indicate that the change in the magnetic exchange energy
is large enough to provide the driving force for high-$T_c$ superconductivity in YBa$_2$Cu$_3$O$_{6.95}$.
}

\begin{figure}[t]
\includegraphics[scale=.35]{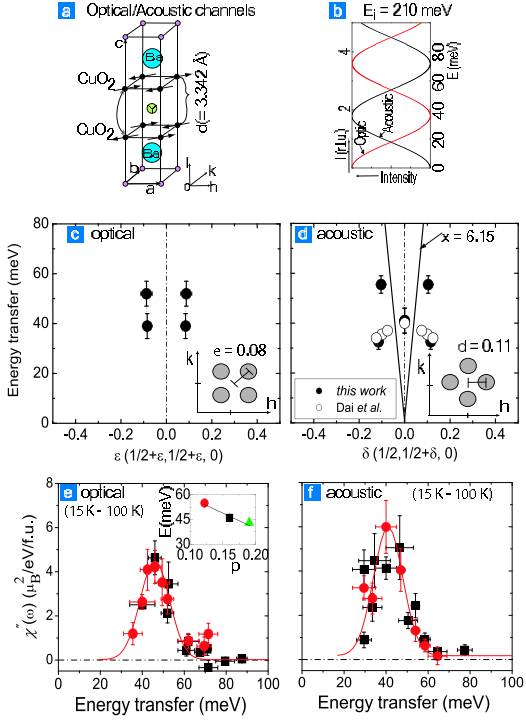}
\caption{Summary of $\mathbf{Q}$- and $\omega$-dependence of
dynamic susceptibility for YBCO.
Our experiments were carried out on the
$\sim$117-g YBCO single crystal ($T_c=92.5$
K) used in previous work \cite{dai01}. We specify the momentum
transfer $(q_{h} ,q_{k}, q_{l})$ (in units of \AA$^{-1}$) as
$(h,k,l) = (q_{h}a/2\pi,q_{k}b/2\pi,q_{l}c/2\pi)$ in reciprocal
lattice units (r.l.u.), where $a = 3.82$, $b=3.88$ and $c
= 11.68$ \AA\ are lattice parameters \cite{dai01}.
(a) Schematic
diagram for YBCO. (b) The $l$-dependence of
acoustic and optical spin fluctuations for $E_i=200$ meV. (c)
Dispersion of constant energy peaks in
$\chi^{\prime\prime}_o(\mathbf{Q},\omega)$. The orientation of
incommensurate peaks is shown in the inset. 
 (d) Dispersion
constant energy peaks in
$\chi^{\prime\prime}_a(\mathbf{Q},\omega)$. The solid line shows
spin-wave dispersion from undoped YBa$_2$Cu$_3$O$_{6.15}$ (Ref.
\cite{hayden96,reznik96}). The inset shows the orientation of the
incommensurate spin fluctuations below the resonance energy.
(e),(f) Local susceptibility of 100 K$-$15 K for optical
$\chi^{\prime\prime}_a(\omega)$ and acoustic
$\chi^{\prime\prime}_o(\omega)$ modes in absolute units. The red
circles and black squares are obtained with cuts along the
$(h,1-h)$ and $(h,h)$ directions with an integrating width of
$\pm$0.15 rlu, respectively. The inset in (e) shows hole-doping
dependence of the optical ``resonance'' with the red circle and
the green triangle from Refs. \cite{pailhes04,pailhes03}. The vertical error bars in (c),(d) and 
the horizontal
error bars in (e),(f) indicate the energy integration range. The vertical
error bars in (e) and (f) are statistical uncertainties (1$\sigma$).}
\end{figure}

If magnetic excitations are mediating electron pairing in the high-$T_c$ copper oxides,
one expects that the change in magnetic exchange energy provides
enough energy for superconducting condensation. The condensation energy
is known experimentally from specific heat measurements for
YBa$_2$Cu$_3$O$_{6.95}$ (YBCO) to be $\sim$3 K/formula unit (f.u.) \cite{loram,Lortz}.
Within the $t$-$J$ model the change in magnetic exchange energy can be
calculated from the nearest neighbor spin correlations
\cite{SW98,DZ98,maier}:

\begin{equation}
 \Delta E_{ex} = 2J \left( \left\langle \vec{S}_i\cdot\vec{S}_{j}\right\rangle_S
-\left\langle\vec{S}_i\cdot\vec{S}_{j} \right\rangle_N \right)
\label{exchangeenergy}
\end{equation}

where $J$ is the exchange interaction, $\vec{S}_i$ and $\vec{S}_j$ are the electron spin
operators at nearest neighbor Cu sites $i$ and $j$ in the CuO$_2$ plane, respectively. 
Instead of estimating the magnetic resonance's contribution to the
superconducting condensation \cite{DZ98,dai99}, we seek here to calculate
$\Delta E_{ex}$ from the entire observable magnetic excitation spectrum. In
general, a complete determination of the magnetic excitation spectrum is
difficult as spin fluctuations can spread over a large wavevector and energy
range. YBCO has two CuO$_2$ planes per unit cell (bilayer) and therefore the
magnetic excitations have odd (acoustic) or even (optical) symmetry with
respect to the neighboring planes (Fig. 1). For optimally doped YBCO, the
magnetic excitation spectrum is dominated by a resonance mode centered at 41
meV in the acoustic channel \cite{mignod,mook}, and a mapping of the acoustic
and optical magnetic excitations should allow an estimation of $\Delta
E_{ex}$.

\begin{figure}[t]
\includegraphics[scale=.35]{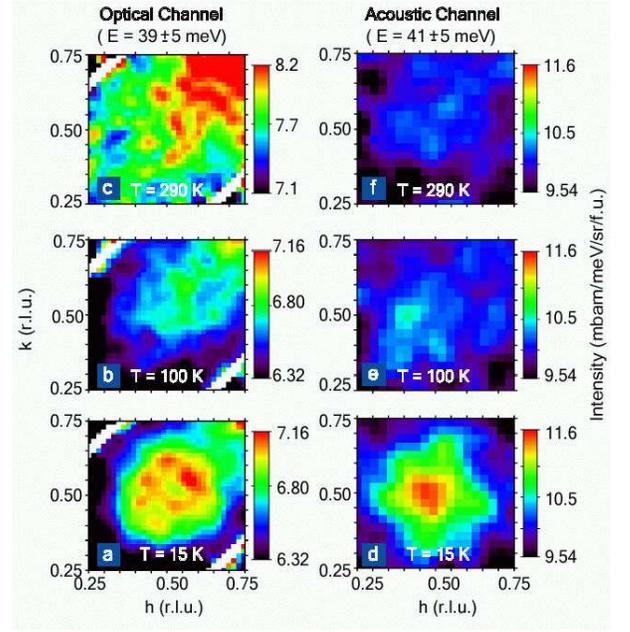}
\caption{ The temperature dependence of the magnetic scattering
around $\hbar\omega\approx 40$ meV at optical (d)-(f), $Ei=90$
meV, and acoustic (a)-(c), $E_i=210$ meV. A clear commensurate
magnetic scattering is seen at $\sim$40 meV at 15 K in (d) while
(a) shows incommensurate scattering. The intensity of phonons
increases with increasing temperature and wavevector (e),(f). 
}
\end{figure}

Figure 1 summarizes the key conclusions of our work. The optical
and acoustic spin fluctuations can be separated by their
differences in $q_z$-dependence (Fig. 1b).  The total magnetic
response $\chi^{\prime\prime}(Q, \omega )$ can then be written as
%%%%SMH
$\chi^{\prime\prime}(q_{x},q_{y},q_{z}, \omega ) =
\chi^{\prime\prime}_{a}(q_{x},q_{y}, \omega )\sin^{2}(q_{z}d/2 ) +
\chi^{\prime\prime}_{o}(q_{x},q_{y}, \omega )\cos^{2}(q_{z}d/2 )$,
%%%%SMH
where $d=3.342$ \AA\ is the spacing between the nearest neighbor
CuO$_2$ planes along the $c$-axis. To probe the entire magnetic
spectra in optical and acoustic channels of
YBCO, we used the MAPS spectrometer at ISIS Facility
\cite{hayden,woo} and chose incident beam energies of $E_i =
30$, 40, 62.5, 75, 90, 110, 130, 138, 160, 210, 280, 360, and 450
meV with the incident beam along the $c$-axis.
The position sensitive detectors on MAPS allow
a complete determination on the $\mathbf{Q}$-structure of
incommensurate spin fluctuations for YBCO in one
experimental setting \cite{hayden}. This avoids the complication of de-convoluting the
instrumental resolution necessary for structure determination of
incommensurate peaks using triple-axis spectroscopy
\cite{reznik04}.
The temperatures probed were $T = 15$, 100, and 290 K. The
intensity difference between 15 K and 100 K is almost
entirely magnetic because of the small value of $Q^2$ probed by the
experiment and small change in the Bose factor for
$\hbar\omega> 30$ meV \cite{fong,dai01,regnault}.

Figures 2(a)-(c) summarize the temperature dependence of the
$34\leq\hbar\omega\leq 44$ meV scattering at the position of
optical spin fluctuations ($E_i=90$ meV). A clear incommensurate
scattering appears at 15 K (Fig. 2(a)) and they are replaced by a
broad response at 100 K (Fig. 2(b)).
Figs. 2(d)-(f) show the
temperature dependence of the scattering around the 41 meV
acoustic resonance obtained by using $E_i=210$ meV. At $T=15$ K,
the scattering shows a sharp resonance centered at $(1/2,1/2)$
\cite{dai01}. On warming to 100 K, the resonance disappears (Fig.
2(e)). Further warming to 290 K does not change the scattering
significantly (Fig. 2(f)).

\begin{figure}[t]
\includegraphics[scale=.35]{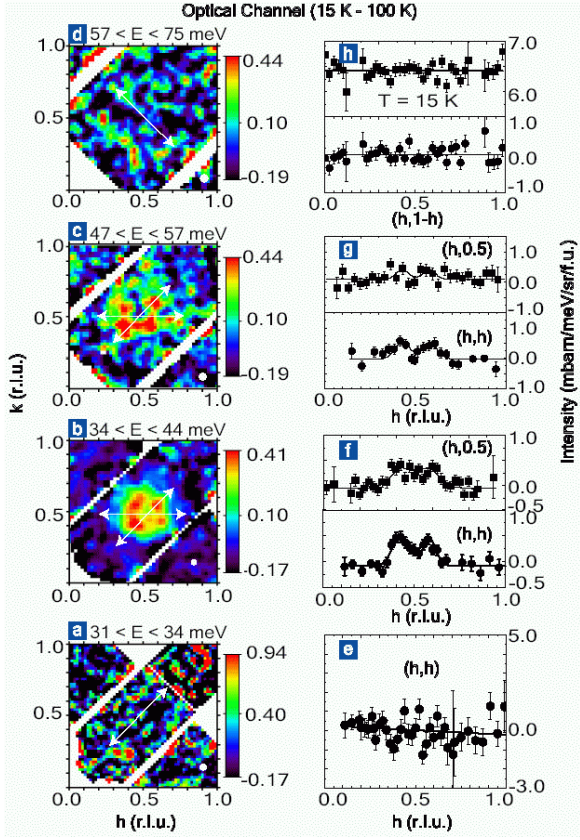}
\caption{ Temperature difference (15 K$-$100 K) at various
energies for optical mode defined as scattering with
$\cos^{2}(q_{z}d/2 )>0.8$. Images (a)-(d) are obtained with
$Ei=75, 90, 130, 210$ meV, respectively. The image at
$\hbar\omega=45.5\pm 1.5$ meV does not have enough statistics to
determine the ${\bf Q}$-structure. Panels (e)-(h) are cuts with
$\hbar\omega=32.5\pm 1.5,39\pm 5,52\pm 5,66\pm 9$ meV,
respectively. The upper panel of (h) shows a cut through the
unsubtracted data of (h) at 15 K. 
The vertical
error bars in (e-h) are statistical uncertainties (1$\sigma$).
}
\end{figure}

Figure 3 summarizes the optical spin fluctuations for
$31\leq\hbar\omega\leq 75$ meV. For $31\leq\hbar\omega\leq 34$
meV, the scattering shows no difference between normal and
superconducting states (Figs. 3(a) and 3(e)). Since there is
little normal state magnetic scattering, there must be an optical
spin gap around 34 meV. On increasing the energy transfer to
$\hbar\omega=39\pm5$ meV, where the acoustic channel has a
commensurate resonance, spin fluctuations in the optical channel
form a broad incommensurate structure away from $(1/2,1/2)$ (Fig.
3(b)). Figure 4f confirms the incommensurate nature of the
scattering and shows that the $(h,h)$ and $(h,0.5)$ cut directions
are inequivalent. For $\hbar\omega=52\pm5$ meV, again we find
incommensurate peaks but this time the scattering is more box-like
with enhanced corners (Figs. 3(c) and 3(g)). The orientation of
the scattering is rotated 45$^\circ$ from that in Fig. 4(b),
similar to acoustic high-energy spin excitations in
YBa$_2$Cu$_3$O$_{6.6}$ \cite{hayden}. Finally, on moving the
energy to $\hbar\omega=66\pm 9$ meV, the temperature difference
spectrum is featureless. Inspection of the unsubtracted data at 15
K reveals no observable magnetic scattering (Fig. 3(h)).

\begin{figure}[t]
\includegraphics[scale=.4]{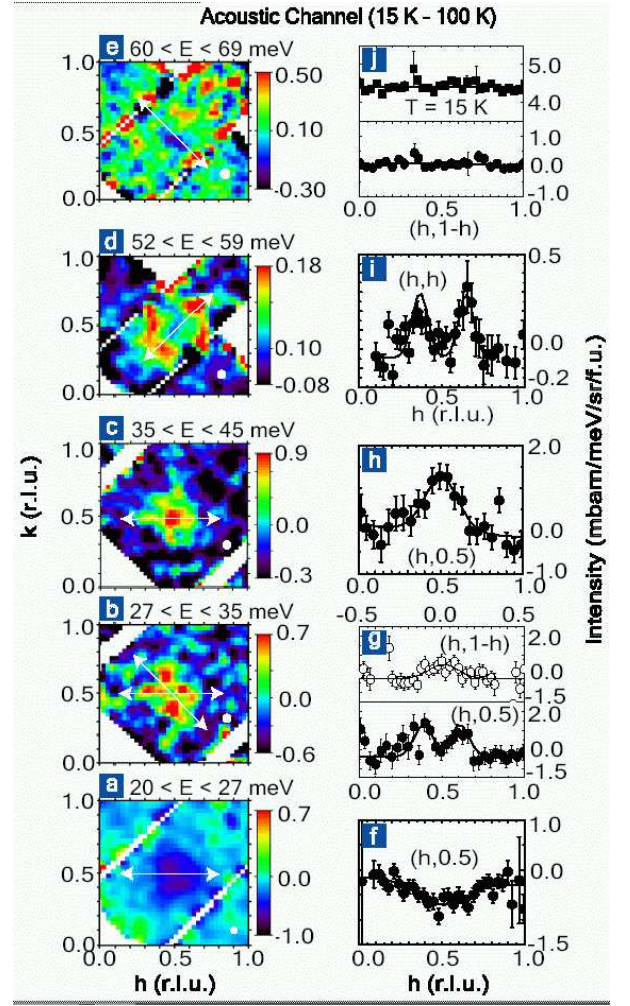}
\caption{ Temperature difference spectra (15 K$-$100 K) at various
energies for acoustic mode defined as any scattering with
$\sin^{2}(q_{z}d/2 )>0.8$ (see Fig. 1d). Images (a)-(e) are
obtained with $Ei=90, 210, 210, 90, 110$ meV, respectively. Cuts
in (f)-(j) are obtained with $\hbar\omega=41\pm 5,31\pm 4,40\pm
5,55.5\pm 3.5,64.5\pm 4.5$ meV, respectively. The upper panel of
(j) shows a cut through the unsubtracted data of (e) at 15 K. The
instrumental $Q$-resolutions are marked by white circles in
(a)-(e). 
The vertical
error bars in (f-j) are statistical uncertainties (1$\sigma$).
}
\end{figure}

Figures 4(a)-(e) show the intensity difference spectra between 15
K and 100 K at various energies in the acoustic channel. For
$20\leq\hbar\omega\leq 27$ meV, the temperature difference has
negative intensity, possibly due to a reduction in the magnetic
response on entering the superconducting state \cite{regnault} or
large phonon population at 100 K (Fig. 4a). A cut through the
image confirms no intensity gain at low temperatures (Fig. 4f).
When increasing the energy to $27\leq\hbar\omega\leq 35$ meV, the
difference image shows a quartet of incommensurate peaks at
$(1/2\pm\delta,1/2)$ and $(1/2,1/2\pm\delta)$ with $\delta=0.11\pm
0.02$ r.l.u. To demonstrate that the incommensurate scattering is
not circular around $(1/2,1/2)$, we made two cuts
through the image. The cut along the $(h,1/2)$ direction clearly
shows two incommensurate peaks around $(1/2,1/2)$. In contrast, a
cut along the $(h,1-h)$ direction has no incommensurate peaks. Note that
a circular symmetry at $\hbar\omega= 35$ meV has been
suggested for YBCO from previous
triple-axis experiments \cite{reznik04}.

Figure 4(c) shows the data at the resonance energy and its
wavevector dependence has a Gaussian lineshape centered around
$(1/2,1/2)$ (Fig. 4(h)). For energies above the resonance
($52\leq\hbar\omega\leq 59$ meV), the scattering is incommensurate
although the low counting rate does not allow an unambiguous
identification of its $\mathbf{Q}$-structure (Figs. 4(d),4(i)).
For $\hbar\omega>60$ meV, the temperature difference spectra as
well as unsubtracted data show no evidence of magnetic scattering
around $(1/2,1/2)$ (Figs. 4(e),4(j)).

Figures 1(e) and 1(f) summarize the superconductivity-induced (15
K$-$100 K) change in the local dynamic susceptibility
$\chi^{\prime\prime}(\omega)$ in absolute units for optical and
acoustic channels of YBCO, respectively. In the
optical channel, $\chi^{\prime\prime}_{o}(\omega)$ has a peak at
$\hbar\omega = 46$ meV, similar to theoretical expectations \cite{wen}. The magnetic spectral weight gradually increases
from above the spin-gap value of $\sim$30 meV \cite{dai01}, peaks
at 46 meV, and finally diminishes for energies above $\sim$70 meV.
The acoustic channel behaves similarly although it peaks at the
expected resonance position of 40 meV \cite{dai01}.
The magnitude of the total spectral weight in the optical channel
$\left\langle m^2\right\rangle_{op}
= 0.078\pm0.02$
$\mu_B^2$  and in the acoustic channel is $\left\langle
m^2\right\rangle_{ac}
= 0.102\pm0.02$
$\mu_B^2$ per f.u. respectively.
This value is similar to that for
YBa$_2$Cu$_3$O$_{6.6}$ around the acoustic resonance energy ($\left\langle
m^2\right\rangle_{ac} = 0.12\pm0.02$ $\mu_B^2$/f.u. for
$24<\hbar\omega<44$
meV) \cite{hayden}. Since the high-energy
response in underdoped YBa$_2$Cu$_3$O$_{6+x}$ ($x=0.5,0.6$)  takes
up much more spectral weight than the resonance
\cite{stock,hayden}, it is surprising that there is essentially no
observed magnetic response for energies above 60 meV in
YBCO (Fig. 1f). Compared to undoped
YBa$_2$Cu$_3$O$_{6.15}$ \cite{hayden96,reznik96}, which has a
total integrated moment of $\sim$0.4 $\mu_B^2$/f.u. when
integrated up to 120 meV, the total integrated moment in optimally
doped YBCO has only about 26\% of the spectral
weight in the same energy range in the acoustic channel only.

Using the spin excitation spectra in Figs. 3 and 4,
we have calculated the changes in the magnetic excitations from the normal to the superconducting
state and estimated
$\delta \left\langle \vec{S}_i\cdot\vec{S}_j\right\rangle=-0.020\pm0.008$/f.u.,
where $\left\langle \vec{S}_i\cdot\vec{S}_j\right\rangle$ is the spin-spin correlation
function for nearest neighbour copper atoms (see supplementary information).
This estimate neglects contributions from energies below 24 meV and
above 70~meV, where
magnetic scattering is difficult to resolve. Also, in Eq.~(\ref{exchangeenergy}) the difference
between normal and superconducting state is meant to be determined at the same
temperature, while here we had to take normal state data at 100~K and
superconducting state data at 15~K neglecting a possible temperature
dependence of the normal state magnetic excitations. In order to
assess the error introduced by these neglections, we have fitted an
RPA-BCS model calculation of the spin excitation spectrum \cite{dahm} to our data
and calculated the missing contributions within this model.
This calculation indicates that our value for $\Delta E_{ex}$ could
be of order 30~\% too large due to these neglections(see supplementary information).

Assuming an exchange coupling of $J=100$ meV, the change in exchange energy
would be $\Delta E_{ex}=2J \; \delta\left\langle \vec{S}_i\cdot\vec{S}_j\right\rangle
=-4.1$ meV/f.u.=-24 K/planar~Cu. This value is a factor of 1.3 times larger
than the 18 K/Cu estimated from the acoustic resonance alone
in previous work \cite{DZ98}. Even if we consider that our estimation may be
too large by  30~\%, the change in the exchange energy is still much larger
than the $U_0 \cong 25 (J/mole) \simeq$ 0.26 meV/f.u.= 3 K/f.u. = 1.5~K/planar~Cu
condensation energy \cite{loram,Lortz}.

Our results reveal two important conclusions for spin
excitations of optimally doped YBCO. First,
the optical resonance reported earlier \cite{pailhes04,pailhes03}
is actually incommensurate and this naturally explains the large
$Q$-widths previously reported. Second, our determination of the
dynamical susceptibility in absolute units allows an estimation of
the change of the magnetic excitation energy available
to the superconducting condensation energy
\cite{DZ98,chakravarty,kivelson}. We find that
the magnetic exchange energy is about 15 times larger than that of
the superconducting condensation energy, thus indicating magnetism
can be the driving force for electron pairing and superconductivity.

%\bibliography{NoEndingPoint}

\begin{thebibliography}{}
\bibitem{Sca95} Scalapino, D. J. The case for $d_{x^2-y^2}$
pairing in the cuprate superconductors. Phys. Reports {\bf 250}, 330-365 (1995).
\bibitem{CPS03} Chubukov, A., Pines, D., \& Schmalian, J., in The Physics of Superconductors,
Vol I, Conventional and High-$T_c$ Superconductors (ed.~by Bennemann, K.H. \& Ketterson, J.B.)
495-590 (Springer, Berlin, 2003).
\bibitem{SW98} Scalapino, D.~J. \& White, S.~R.
Superconducting condensation energy and an antiferromagnetic exchange-based
pairing mechanism. Phys. Rev. B {\bf 58}, 8222-8224 (1998).
\bibitem{DZ98} Demler, E. \& Zhang, S-C.
Quantitative test of a microscopic mechanism of high-temperature superconductivity.
Nature {\bf 396}, 733-735 (1998).
\bibitem{maier} Maier, Th. A. On the nature of pairing in the two-dimensional $t$-$J$ model.
Physica B {\bf 359-361}, 512-514 (2005). 
\bibitem{mignod} Rossat-Mignod, J. {\it et al.} Neutron scattering study of the
YBa$_2$Cu$_3$O$_{6+x}$ system. Physica C {\bf 185}, 86-92 (1991).
\bibitem{mook} Mook, H. A. {\it et al.} Polarized neutron determination of the
magnetic excitations in YBa$_2$Cu$_3$O$_{7}$. Phys. Rev. Lett. {\bf 70}, 3490-3493 (1993).
\bibitem{mook98} Mook, H. A. {\it et al.} Spin fluctuations in YBa$_2$Cu$_3$O$_{6.6}$.
Nature {\bf 395}, 580-582 (1998).
\bibitem{fong} Fong, H. F. {\it et al.} Spin susceptibility in underdoped
YBa$_2$Cu$_3$O$_{6+x}$. Phys. Rev. B {\bf 61}, 14773-14786 (2000).
\bibitem{dai01} Dai, P., Mook, H. A., Hunt, R. D., \& Do$\rm\breve{g}$an, F.
Evolution of the resonance and incommensurate spin fluctuations in superconducting YBa$_2$Cu$_3$O$_{6+x}$.
 Phys. Rev. B {\bf 63}, 054525 (2001).
\bibitem{stock} Stock, C. {\it et al.} From incommensurate to dispersive spin-fluctuations: The high-energy inelastic
spectrum in superconducting YBa$_2$Cu$_3$O$_{6.5}$.
Phys. Rev. B {\bf 71}, 024522 (2005).
\bibitem{hayden} Hayden, S. M., Mook, H. A., Dai, P., Perring, T. G., \& Do$\rm\breve{g}$an, F.
The structure of the high-energy spin excitations in a high-transition-temperature
superconductor. Nature {\bf 429}, 531-534 (2004).
\bibitem{dai99} Dai, P. {\it et al.}, The magnetic excitation spectrum and thermodynamics
of high-$T_c$ superconductors. Science {\bf 284}, 1344-1347 (1999).
\bibitem{kivelson} Kee, H-Y., Kivelson, S. A. \&
Aeppli, G.
Spin-1 Neutron Resonance Peak Cannot Account for Electronic Anomalies in the Cuprate Superconductors.
 Phys. Rev. Lett. {\bf 88}, 257002 (2002).
\bibitem{loram} Loram, J. W., Mirza, K. A., Cooper, J. R. \& Tallon, J. L. Specific heat
evidence on the normal state pseudogap, J. Phys. Chem. Solids. {\bf 59}, 2091-2094 (1998).
\bibitem{Lortz} Lortz, R. {\it et al.}
Evolution of the specific-heat anomaly of the high-temperature
superconductor in YBa$_2$Cu$_3$O$_7$ under the influence of doping
through application of pressure up to 10 GPa. J.Phys.: Condens.
Matter {\bf 17}, 4135-4145 (2005).
\bibitem{woo} Woo, H. {\it et al.}
Mapping spin-wave dispersions in stripe-ordered La$_{2–x}$Sr$_x$NiO$_4$ ($x=0.275, 0.333$).
Phys. Rev. B {\bf 72}, 064437 (2005).
\bibitem{reznik04} Reznik, D. {\it et al.}
Dispersion of Magnetic Excitations in Optimally Doped Superconducting YBa$_2$Cu$_3$O$_{6.95}$.
Phys. Rev. Lett. {\bf 93}, 207003 (2004).
\bibitem{regnault} Regnault, L. P. {\it et al.} Spin dynamics in the high-$T_c$ superconducting system
YBa$_2$Cu$_3$O$_{6+x}$. Physica B {\bf 213 \& 214}, 48-53 (1995).
\bibitem{hayden96} Hayden, S. M. {\it et al.}
High-frequency spin waves in YBa$_2$Cu$_3$O$_{6.15}$.
Phys. Rev. B {\bf 54}, R6905-R6908 (1996).
\bibitem{reznik96} Reznik, D. {\it et al.}
Direct observation of optical magnons in YBa$_2$Cu$_3$O$_{6.2}$.
Phys. Rev. B {\bf 53}, R14741-R14744 (1996).
\bibitem{dahm}
Dahm, T. {\it et al.}, Nodal quasiparticle lifetimes in
cuprate superconductors.  Phys. Rev. B {\bf 72}, 214512  (2005).
\bibitem{pailhes04} Pailhes, S. {\it et al.}
Resonant Magnetic Excitations at High Energy in Superconducting YBa$_2$Cu$_3$O$_{6.85}$.
Phys. Rev. Lett. {\bf 93}, 167001 (2004).
\bibitem{pailhes03} Pailhes, S. {\it et al.}
Two Resonant Magnetic Modes in an Overdoped High $T_c$ Superconductor.
Phys. Rev. Lett. {\bf 91}, 237002 (2003).
\bibitem{wen} Chen, W. Q. \& Weng, Z. Y. Spin dynamics in a 
doped-Mott-insulator superconductor. Phys. Rev. B {\bf 71}, 134516 (2005).
\bibitem{chakravarty} Chakravarty, S. \& Kee, H-Y.
Measuring condensate fraction in superconductors.
Phys. Rev. B {\bf 61}, 14821-14824 (2000).

\end{thebibliography}

Correspondence and requests for materials should be addressed to P.D. (daip@ornl.gov) or H.A.M. (ham@ornl.gov).

%\acknowledgments
\begin{acknowledgments}
{\bf Acknowledgments} We thank Elbio Dagotto, Z. Y. Wen, and F. C. Zhang for helpful discussions. 
This work is supported by the US DOE Office of Science, Division of Materials Science, Basic
Energy Sciences under contract No. DE-FG02-05ER46202 (H.W. and P.D.).
Oak Ridge National Laboratory is supported by the US DOE
under contract No. DE-AC05-00OR22725
with UT/Battelle LLC.  SMH is supported by the UK EPSRC.  DJS would like to acknowledge the Center for
Nanophase Material Science at Oak Ridge National Laboratory for their support.
Supplementary Information accompanies this paper on www.nature.com/naturephysics.
\end{acknowledgments}

{\bf Competing financial interests}

The authors declare that they have no competing financial interests.

\end{document}